\documentclass{emulateapj} 

\gdef\rfUV{(U-V)_{rest}}
\gdef\rfVJ{(V-J)_{rest}}
\gdef\1054{MS\,1054--03}

\def\farcs{\hbox{$.\!\!^{\prime\prime}$}}
\def\simgeq{{\raise.0ex\hbox{$\mathchar"013E$}\mkern-14mu\lower1.2ex\hbox{$\mathchar"0218$}}} 

\begin {document}

\title {What do we learn from IRAC observations of galaxies at $2 < z < 3.5$?\altaffilmark{1}}

\author{Stijn Wuyts\altaffilmark{2}, Ivo Labb\'{e}\altaffilmark{3,4}, Marijn Franx\altaffilmark{2}, Gregory Rudnick\altaffilmark{5}, Pieter G. van Dokkum\altaffilmark{6}, Giovanni G. Fazio\altaffilmark{7}, Natascha M. F\"{o}rster Schreiber\altaffilmark{8}, Jiasheng Huang\altaffilmark{7}, Alan F. M. Moorwood\altaffilmark{9}, Hans-Walter Rix\altaffilmark{10}, Huub R\"{o}ttgering\altaffilmark{2}, Paul van der Werf\altaffilmark{2}}
\altaffiltext{1}{Based on observations with the {\it Spitzer Space Telescope}, which is operated by the Jet Propulsion Laboratory, California Institute of Technology under NASA contract 1407.  Support for this work was provided by NASA through contract 125790 issued by JPL/Caltech.  Based on service mode observations collected at the European Southern Observatory, Paranal, Chile (LP Program 164.O-0612).  Based on observations with the NASA/ESA {\it Hubble Space Telescope}, obtained at the Space Telescope Science Institute which is operated by AURA, Inc., under NASA contract NAS5-26555.}
\altaffiltext{2}{Leiden Observatory, P.O. Box 9513, NL-2300 RA, Leiden, The Netherlands [e-mail: wuyts@strw.leidenuniv.nl]}
\altaffiltext{3}{Carnegie Observatories, 813 Santa Barbara Street, Pasadena, CA 91101}
\altaffiltext{4}{Carnegie Fellow}
\altaffiltext{5}{National Optical Astronomical Observatory, 950 North Cherry Avenue, Tucson, AZ 85721}
\altaffiltext{6}{Department of Astronomy, Yale University, New Haven, CT 06520-8101}
\altaffiltext{7}{CFA, 60 Garden Street Cambridge, MA 02138}
\altaffiltext{8}{MPE, Giessenbackstrasse, D-85748, Garching, Germany}
\altaffiltext{9}{ESO, D-85748, Garching, Germany}
\altaffiltext{10}{MPIA, K\"{o}nigstuhl 17, D-69117, Heidelberg, Germany}

\begin{abstract}
We analyze very deep HST, VLT and Spitzer photometry of galaxies at
$2<z<3.5$ in the Hubble Deep Field South.  The sample is selected from
the deepest public K-band imaging currently available.  We show that
the rest-frame $U-V$ vs $V-J$ color-color diagram is a powerful
diagnostic of the stellar populations of distant galaxies.  Galaxies
with red rest-frame $U-V$ colors are generally red in rest-frame $V-J$
as well.  However, at a given $U-V$ color a range in $V-J$ colors
exists, and we show that this allows us to distinguish young, dusty
galaxies from old, passively evolving galaxies.  We quantify the
effects of IRAC photometry on estimates of masses, ages, and the dust
content of $z>2$ galaxies.  The estimated distributions of these
properties do not change significantly when adding IRAC data to the
UBVIJHK photometry.  However, for individual galaxies the addition of
IRAC can improve the constraints on the stellar populations,
especially for red galaxies: uncertainties in stellar mass decrease by
a factor of 2.7 for red ($\rfUV > 1$) galaxies, but only by a factor
of 1.3 for blue ($\rfUV < 1$) galaxies.  We find a similar
color-dependence of the improvement for estimates of age and dust
extinction.  In addition, the improvement from adding IRAC depends on
the availability of full near-infrared JHK coverage; if only K-band
were available, the mass uncertainties of blue galaxies would decrease
by a more substantial factor 1.9.  Finally, we find that a trend of
galaxy color with stellar mass is already present at $z>2$.  The most
massive galaxies at high redshift have red rest-frame $U-V$ colors
compared to lower mass galaxies even when allowing for complex star
formation histories.
\end{abstract}


\section {Introduction}
\label{intro.sec}

Two of the major challenges in observational cosmology are
understanding the history of star formation in galaxies, and
understanding the assembly of mass through cosmic time.  In the local
universe elaborate surveys mapped the diversity of nearby galaxies
(e.g. Blanton et al. 2003) and characterized the dependence of their
colors (Baldry et al. 2004) and star formation (Kauffmann et al. 2003)
on galaxy mass.  The study of their progenitors at $z \simgeq 2$ is
important since it is believed that at this epoch the most massive
galaxies formed their stars (Glazebrook et al. 2004; van der Wel et
al. 2005; Rudnick et al. 2006).

The first method to efficiently identify distant galaxies was the
Lyman-break technique (Steidel et al. 1996).  Large samples have been
spectroscopically confirmed (Steidel et al. 1999).  Their stellar
populations have been characterized by means of broad-band photometry
(e.g. Papovich, Dickinson, \& Ferguson 2001, Shapley et al. 2005),
optical spectroscopy (e.g. Shapley et al. 2003) and near-infrared
spectroscopy (Erb et al. 2003, 2006).  Lyman break galaxies (LBGs)
have spectral energy distributions similar to nearby starburst
galaxies.

In recent years, new selection criteria provided evidence for a
variety in color space among high redshift galaxies as rich as in the
local universe.  Among the newly discovered populations are submm
galaxies (e.g. Smail et al. 2004), ``IRAC Extremely Red Objects''
(IEROs; Yan et al. 2004), ``BzK'' objects (Daddi et al. 2004) and
distant red galaxies (DRGs; Franx et al. 2003).  The latter are
selected by the simple color criterion $(J-K)_{Vega}>2.3$.  Their
rest-frame UV-to-optical SEDs resemble those of normal nearby galaxies
of type Sbc-Scd (F\"{o}rster Schreiber et al. 2004).  Near-infrared
spectroscopy of DRGs (Kriek et al. 2006) and extension of the
broad-band photometry to mid-infrared wavelengths (Labb\'{e} et
al. 2005) suggests that evolved stellar populations exist already at
$2<z<3.5$.  Rudnick et al. (2006) showed that DRGs contribute
significantly to the mass density in rest-frame optically luminous
galaxies.  van Dokkum et al. (2006) studied a stellar mass-limited
sample of galaxies with $M > 10^{11}\ M_{\sun}$ and found that DRGs,
rather than LBGs, are the dominant population at the high mass end at
$2<z<3$.

In this paper, we exploit the $3-8\mu m$ imaging of the Hubble Deep
Field South by {\it Spitzer}/IRAC (Fazio et al. 2004) to extend the
SED analysis of distant galaxies to the rest-frame near-infrared and
constrain their stellar masses and stellar populations.  Our sample is
complete to $K_{tot,AB} = 25$.  No color selection criteria are applied.
The depth of our imaging allows us to probe down to stellar masses of
a few $10^9\ M_{\sun}$.  We investigate whether IRAC helps to study
the diversity of galaxies at high redshift and if the addition of IRAC
improves the constraints on stellar mass, age and dust content.
Finally, we investigate the dependence of galaxy color on stellar
mass.

The paper is structured as follows.  In \S\ref{Data_phot_sample.sec}
we describe the data, IRAC photometry and sample definition.
\S\ref{SEDmodeling.sec} explains the modeling of spectral energy
distributions (SEDs).  The rest-frame optical to near-infrared color
distribution of our K-selected sample is discussed in
\S\ref{color_distribution.sec}.  \S\ref{constraints.sec} provides an
in depth discussion of the constraints that IRAC places on estimates
of age, dust extinction and stellar mass.  First wavelength and model
dependence are discussed from a theoretical perspective.  Next we
discuss results from applying the models to our U-to-$8\mu m$ spectral
energy distributions.  In \S\ref{mass_rfUV.sec} we investigate the
rest-frame optical colors of high redshift galaxies as a function of
stellar mass.  Finally, the conclusions are summarized in
\S\ref{conclusions.sec}.

Throughout the paper we adopt a cosmology with $H_0 = 70\ km\ s^{-1}\
Mpc^{-1}$, $\Omega_m = 0.3$, and $\Omega_{\Lambda} = 0.7$.

\section {Data, photometry and sample selection}
\label{Data_phot_sample.sec}
\subsection {Data}
\label{data.sec}
Observations of the HDFS/WFPC2 field were obtained with the IRAC
camera (Fazio et al. 2004) on the {\it Spitzer Space Telescope}
(Werner et al. 2004) in June 2004 and June 2005 (GTO program 214).  A
$5' \times 5'$ field of view was covered by the 4 broadband filters at
3.6, 4.5, 5.8 and 8 microns.  The data, reduction and photometry will
be described in detail by Labb\'{e} et al. (in preparation).  Briefly,
we started with the Basic Calibrated Data (BCD) as provided by the
Spitzer Science Center pipeline.  We applied a series of procedures to
reject cosmic rays and remove artifacts such as column pulldown,
muxbleed, and the ``first frame effect'' (Hora et al. 2004).  Finally,
the frames were registered to and projected on a 2x2 blocked
(0\farcs2396 pixel scale) version of an existing ISAAC K-band image
(Labb\'{e} et al. 2003, hereafter L03)\footnote[1]{NIR data from the
FIRES survey of the HDFS is publicly available from
http://www.strw.leidenuniv.nl/\textasciitilde fires}, and
average-combined.  Characteristics such as exposure time, FWHM,
limiting depth ($5\sigma$, 3'' diameter aperture) and positional
accuracy in each of the 4 IRAC bands are summarized in Table\
\ref{data.tab}.  A summary of the optical-to-NIR observations by L03
is provided in Table\ \ref{L03.tab}.  All magnitudes quoted in this
paper are in the AB system.

\begin{deluxetable*}{ccccc}
\tablecolumns{5}
\tablewidth{0pc}
\tablecaption{Characteristics of the IRAC observations \label{data.tab}
}
\tablehead{
\colhead{Filter} & \colhead{Exposure time} & \colhead{FWHM} & \colhead{Limiting depth} & \colhead{Positional Accuracy\tablenotemark{a}} \\
\colhead{($\mu$)} & \colhead{(hr)} & \colhead{($''$)} & \colhead{($5\sigma$, 3$''$ diameter aperture)} & \colhead{($''$)}
}
\startdata
3.6   &   3.76  &  1.95  &  25.6  &  0.09 \\
4.5   &   3.76  &  1.90  &  25.6  &  0.15 \\
5.8   &   3.76  &  2.10  &  23.4  &  0.14 \\
8.0   &   3.64  &  2.15  &  23.3  &  0.11
\enddata
\tablenotetext{a}{The rms difference between bright star positions in IRAC and K-band image.}
\end{deluxetable*}

\begin{deluxetable*}{ccccc}
\tablecolumns{5}
\tablewidth{0pc}
\tablecaption{Characteristics of the optical-to-NIR observations (see L03) \label{L03.tab}
}
\tablehead{
\colhead{Instrument/Telescope} & \colhead{Filter} & \colhead{Exposure time} & \colhead{FWHM} & \colhead{Limiting depth} \\
\colhead{} & \colhead{} & \colhead{(hr)} & \colhead{($''$)} & \colhead{($5\sigma$, $0\farcs7$ diameter aperture)}
}
\startdata
WFPC2/HST   &  $F300W$  &  36.8  &  0.16  & 27.8  \\
WFPC2/HST   &  $F450W$  &  28.3  &  0.14  & 28.6  \\
WFPC2/HST   &  $F606W$  &  27.0  &  0.13  & 28.9  \\
WFPC2/HST   &  $F814W$  &  31.2  &  0.14  & 28.3  \\
ISAAC/VLT   &  $J_s$   &  33.6  &  0.45  & 26.9  \\
ISAAC/VLT   &  $H$     &  32.3  &  0.48  & 26.4  \\
ISAAC/VLT   &  $K_s$   &  35.6  &  0.46  & 26.4
\enddata
\end{deluxetable*}

\subsection {Photometry}
\label{photometry.sec}
\begin {figure} [tp]
\centering
\resizebox{\hsize}{!}{\plotone{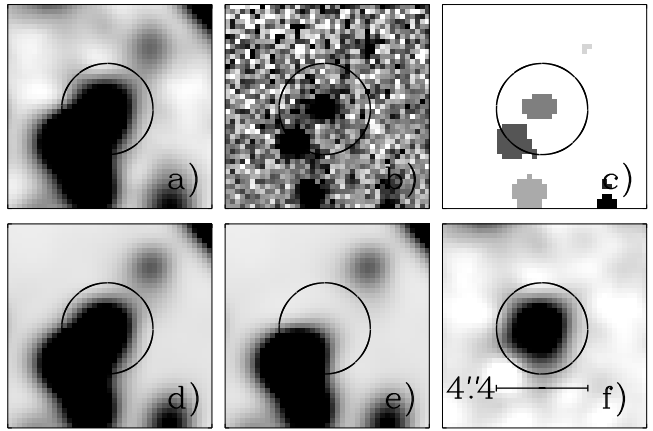}}
\caption{Postage stamps ($9\farcs8 \times 9\farcs8$) illustrating the
deblending procedure for IRAC photometry.  Confusion by nearby
neighbors in the original $3.6\mu m$ image (a) is reduced using the
higher resolution K-band image (b) and its SExtractor segmentation map
(c).  A model $3.6\mu m$ image (d) is created using information on
position and extent of the galaxies from the K-band image.  The model
of the nearby neighbors (e) is subtracted from the original image to
obtain a cleaned $3.6\mu m$ image (f).
\label {method.fig}
}
\end {figure}

In this section we describe the steps to combine the IRAC data and
optical-to-NIR data (L03) into one consistent K-band selected
photometric catalog.  In this paper we limit ourselves to the $2.5'
\times 2.5'$ field where very deep K-band data is available from L03.
The main challenge in doing IRAC photometry is a proper treatment of
source confusion and PSF matching of the data.  Integrating for nearly
4 hours with IRAC at 3.6$\mu m$ and 4.5$\mu m$ reaches a depth only 1
mag shallower than 36 hours of ISAAC K-band imaging ($10\sigma\ limit\
K_{tot,AB}=25$), but the IRAC images have a 4 times broader PSF
causing many sources to be blended.  Information on the position and
extent of K-band detected objects was used to fit and subtract the
fluxes of neighbouring sources.  Each K-band detected source was
isolated using the SExtractor ``segmentation map'' and convolved
individually to the considered IRAC PSF.  Next all convolved sources
were fitted to the IRAC image, leaving only their fluxes as free
parameters.  We subsequently subtract the best-fit fluxes of all
neighboring sources to remove the contamination.  An illustration of
this measurement technique is presented in Fig\ \ref{method.fig}.
The resulting cleaned IRAC images are matched to the broadest PSF (of
the 8$\mu m$ image).  We measured fluxes on the cleaned, PSF-matched
images within a fixed $4\farcs4$ diameter circular aperture.  The
aperture size is a compromise between quality of PSF matching (within
3\% as derived from dividing growthcurves) and adding too much noise.
Finally we applied for each source an aperture correction to scale the
IRAC fluxes to the ``color'' apertures defined for the K-band catalog
by L03.  The correction factor is the ratio of the original K-band
flux in the color aperture and the K-band flux in the $8\mu m$ PSF
matched image within a $4\farcs4$ diameter aperture.  Photometric
errors were calculated by taking the rms of fluxes in $4\farcs4$
diameter apertures on empty places in the IRAC image.  The end product
is a photometric catalog with consistent photometry from optical to
MIR wavelengths with 11 filters (UBVIJHK+IRAC).

\subsection {Sample selection}
\label{sample.sec}
From the catalog described in \S\ref{photometry.sec} we selected all
galaxies, well covered by all 11 filters, that have $S/N > 10$ in the
K-band.  The sample reaches to a limiting total K-band magnitude
$K_{tot, AB} = 25$.

Since spectroscopic redshifts are only available for 63 out of 274
objects, we mostly rely on photometric redshift estimates to select
high redshift galaxies and compute rest-frame colors and luminosities.
The photometric redshifts and derived rest-frame photometry were
calculated as follows.  We used an algorithm developed by Rudnick et
al. (2001, 2003) to fit a nonnegative linear combination of galaxy
templates to the spectral energy distribution of each galaxy.  The
template set consisted of empirical E, Sbc, Scd and Im templates from
Coleman, Wu,\& Weedman (1980), the two least reddened starburst
templates from Kinney et al. (1996) and two Bruzual \& Charlot (2003;
BC03) single stellar populations (SSP) with a Salpeter (1955) stellar
initial mass function (IMF), aged 1 Gyr and 10 Myr respectively.
The empirical templates were extended into the IR using the BC03
stellar population synthesis code.  The derived photometric redshifts
show a good agreement with the available spectroscopic redshifts.  The
average value of $|z_{spec} - z_{phot}|/(1+z_{spec})$ is 0.06, 0.09
and 0.08 for galaxies at $0<z<1$, $1<z<2$ and $2<z<3.5$ respectively.

Once the redshift was derived, we calculated rest-frame luminosities
and colors by interpolating between observed bands using the best-fit
templates as a guide.  For a detailed description, we refer to Rudnick
et al. (2003).

The K-band selected sample contains 121 sources at $0<z<1$, 72 at
$1<z<2$ and 75 at $2<z<3.5$.  The K+IRAC photometry of the galaxies at
$2<z<3.5$ is provided as an electronic table (see Table\
\ref{phot_z2_35.tab} for a sample entry).  In
\S\ref{color_distribution.sec} we study the color-distribution of
galaxies with $L_V > 5 \times 10^9 L_{\sun}$ over the whole redshift
range.  From thereon, we will focus on the high redshift bin.  Two
commonly color-selected populations at $z>2$ will be highlighted where
of interest.  Lyman-break galaxies (LBGs) are selected from the WFPC2
imaging using the criteria of Madau et al. (1996).  Distant Red
Galaxies (DRGs) are selected by the simple color criterion
$(J-K)_{AB} > 1.34$ (Franx et al. 2003).
\begin{deluxetable*}{lrrrrrr}
\tablecolumns{5}
\tablewidth{0pc}
\tablecaption{K+IRAC photometry of HDFS galaxies at $2<z<3.5$ [example entry of electronic table] \label{phot_z2_35.tab}
}
\tablehead{
\colhead{Object\tablenotemark{a}} & \colhead{$f_{K,tot}$\tablenotemark{b}} & \colhead{$f_{K,col}$} & \colhead{$f_{3.6\mu m, col}$} & \colhead{$f_{4.5\mu m, col}$} & \colhead{$f_{5.8\mu m, col}$} & \colhead{$f_{8.0\mu m, col}$} \\
}
\startdata
176  &  $  6.75 \pm   0.47$  &  $  5.07 \pm   0.17$  &  $  8.60 \pm   0.23$  &  $ 10.68 \pm   0.25$  &  $ 14.43 \pm   1.74$  &  $  9.30 \pm   2.06$
\enddata
\tablenotetext{a}{Object identification number corresponds to that of the U-to-K catalog by Labb\'{e} et al. (2003).}
\tablenotetext{b}{Fluxes in total (tot) and color (col) aperture are scaled to an AB zeropoint of 25, i.e. mag$_{AB} = 25 - 2.5 \log f$.}
\end{deluxetable*}
\begin{deluxetable*}{lrrrrr}
\tablecolumns{5}
\tablewidth{0pc}
\tablecaption{Modeling results for HDFS galaxies at $2<z<3.5$ [example entry of electronic table] \label{model_z2_35.tab}
}
\tablehead{
\colhead{Object\tablenotemark{a}} & \colhead{z} & \colhead{SFH} & \colhead{$\log(M_{*})$} & \colhead{$A_V$} & \colhead{$\log(Age_w)$} \\
\colhead{} & \colhead{} & \colhead{} & \colhead{($M_{\sun}$)} & \colhead{} & \colhead{(Gyr)} \\
}
\startdata
176 & $  2.76 ^{+  0.10} _{-  0.34}$  &  $\tau_{300}$  & $  10.77 ^{+  0.07} _{-  0.04}$  &  $    0.6 ^{+   0.4} _{-   0.2}$  &   $ -0.25 ^{+  0.26} _{-  0.00}$
\enddata
\tablenotetext{a}{Object identification number corresponds to that of the U-to-K catalog by Labb\'{e} et al. (2003).}
\end{deluxetable*}

\section {SED modeling}
\label{SEDmodeling.sec}

\begin {figure*} [tp]
\centering
\plotone{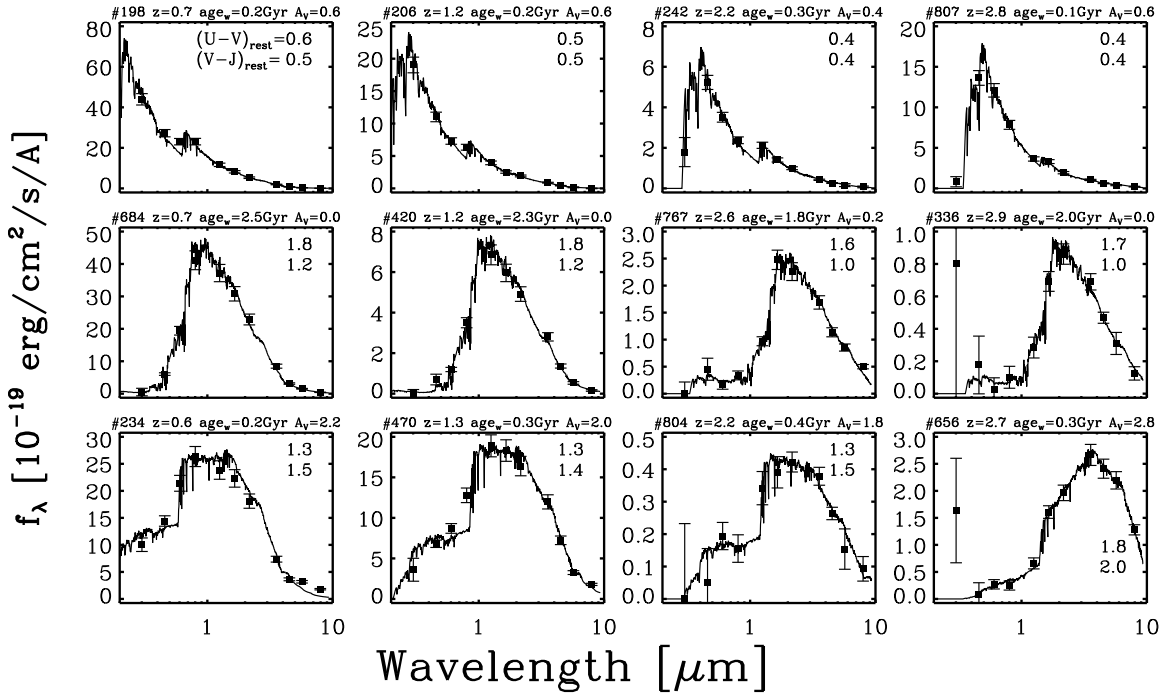}
\caption{The U-to-$8\mu m$ spectral energy distributions of a subset
of galaxies occupying different locations in $\rfUV$ vs $\rfVJ$
color-color space.  Each row shows observed and BC03 model SEDs for
galaxies with redshifts ranging from $z \sim 0.7$ to $z \sim 3$.  A
broad range of galaxy types is present at all redshifts.  Galaxies
with blue $\rfUV$ colors (upper row) have young ages and a modest
amount of dust obscuration.  Objects with red $\rfUV$ colors that are
on the blue side of the $\rfVJ$ color distribution (middle row) are
best fit by old stellar populations with little dust obscuration.  The
bottom row shows examples of galaxies with red optical and red
optical-to-NIR colors.  They are consistent with young stellar
populations with a large dust reddening.
\label {SEDfig1.fig}
}
\end {figure*}

To study physical characteristics of the galaxies such as stellar
mass, stellar age and amount of dust extinction, we make use of the
evolutionary synthesis code developed by BC03.  We fitted the
synthetic spectra to our observed SEDs using the publicly available
HYPERZ stellar population fitting code, version 1.1 (Bolzonella et
al. 2000).  Redshifts were fixed to the $z_{phot}$ measurement (see
\S\ref{sample.sec}, Rudnick et al. 2003) or $z_{spec}$ when available.
A minimum error of 0.08 mag was adopted to avoid that the data points
with the largest errors are effectively ignored in the SED fits.  We
fitted three distinct star formation histories: a single stellar
population (SSP) without dust, a constant star-formation (CSF) history
with dust ($A_V$ varying from 0 to 4 in steps of 0.2) and an
exponentially declining star-formation history with an e-folding
timescale of 300 Myr ($\tau_{300}$) and identical range of $A_V$
values.  The exponentially declining model allows for quiescent
systems that underwent a period of enhanced star formation in their
past.  F\"{o}rster Schreiber et al. (2004) showed that the estimated
extinction values do not vary monotonically with the e-folding
timescale $\tau$, but reach a minimum around 300 Myr.  Including the
$\tau_{300}$ model thus ensures that the allowed star formation
histories encompass the whole region of parameter space that would be
occupied when fitting models with different values of $\tau$.  For
each of the star-formation histories (SFHs), we constrained the time
elapsed since the onset of star formation to a minimum of 50 Myr,
avoiding fit results with improbable young ages.  The age of the
universe at the observed redshift was set as an upper limit to the
ages.  Furthermore, we assume a Salpeter (1955) IMF with lower and
upper mass cut-offs $0.1 M_{\sun}$ and $100 M_{\sun}$, solar
metallicity and we adopt a Calzetti et al. (2000) extinction law.  For
each object the star-formation history resulting in the lowest
$\chi^2$ of the fit was selected and corresponding model quantities
such as age, mass and dust extinction were adopted as the best-fit
value.  We calculated the mass-weighted age for each galaxy by
integrating over the different ages of SSPs that build up the SFH,
weighting with their mass fraction.  We use this measure since it is
more robust with respect to degeneracies in SFH than the time passed
since the onset of star formation; it describes the age of the bulk of
the stars.  An electronic table, the contents of which are illustrated
in Table\ \ref{model_z2_35.tab}, summarizing the results of our SED
modeling for the subsample of galaxies at $2<z<3.5$ is available in
the online Journal.  In Fig.\ \ref{SEDfig1.fig} we show example
U-to-$8\mu m$ SEDs with best-fit BC03 models of galaxies over the
whole redshift range, illustrating that at all epochs a large variety
of galaxy types is present.

We fitted all objects in our sample twice, once with and once without
IRAC photometry.  We repeated the SED modeling with the same parameter
settings using the models by Maraston et al. (2005).  The results will
be discussed in \S\ref{constr_obs_mod.sec}.  Variations in modeled
parameters due to a different metallicity are addressed in
\S\ref{metal.sec}.  The effects of adopting a different extinction law
are discussed in \S\ref{extinction.sec}.  Unless mentioned otherwise,
we refer to stellar mass, mass-weighted age and dust extinction values
derived from the U-to-$8\mu m$ SEDs with BC03 models.

\section {Rest-frame optical to near-infrared color distribution}
\label{color_distribution.sec}
At redshifts above 1 all rest-frame near-infrared bands have shifted
redward of observed K and mid-infrared photometry is needed to compute
rest-frame NIR fluxes from interpolation between observed bands.  It
has only been with the advent of IRAC on the {\it Spitzer Space
Telescope} that the rest-frame near-infrared opened up for the study
of high redshift galaxies.  As the $3.6\mu m$ and $4.5\mu m$ images
are much deeper than the $5.8\mu m$ and $8.0\mu m$ images (see Table\
\ref{data.tab}), we focus on the rest-frame J-band ($J_{rest}$).

Several studies have focussed on the optical to near-infrared colors
and inferred stellar populations of particular color-selected samples
(e.g. Shapley et al. 2005, Labb\'{e} et al. 2005).  In this section we
take advantage of the multiwavelength data and the very deep K-band
selection to study the rest-frame optical to near-infrared colors of
all galaxies up to $z= 3.5$ without color bias.  For the first time we
can therefore investigate what range in optical to near-infrared
colors high redshift galaxies occupy, how their optical to
near-infrared colors relate to pure optical colors, and what this
tells us about the nature of their stellar populations.

\begin {figure} [bp]
\centering
\resizebox{\hsize}{!}{\plotone{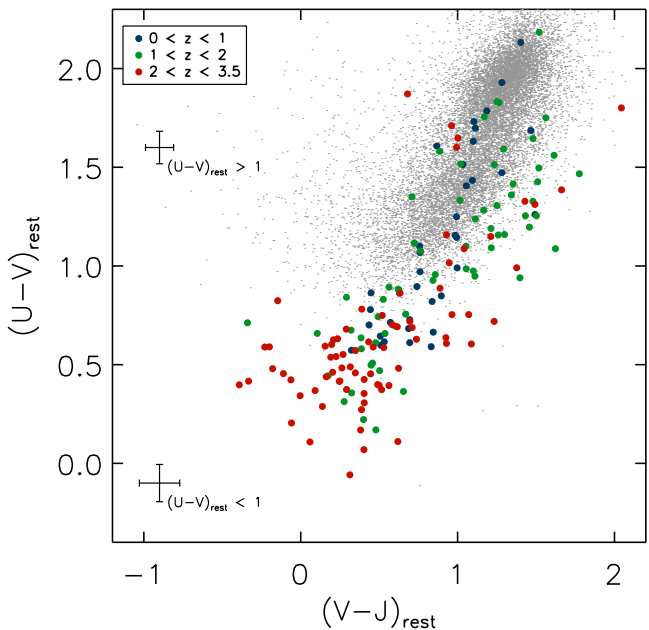}}
\caption{Rest-Frame $U-V$ versus $V-J$ color-color diagram of all
galaxies with $L_V > 5 \times 10^{9}L_{\sun}$.  SDSS+2MASS galaxies
(small grey dots) are plotted as a local reference.  Color coding
refers to the redshift bin.  Galaxies with red $U-V$ colors are also
red in $V-J$.  Compared to the local SDSS galaxies the high redshift
color distribution extends to bluer $U-V$ colors (where Lyman-break
galaxies are located) and for the same $U-V$ color to redder $V-J$
colors.
\label {fig1a.fig}
}
\end {figure}

In Fig.\ \ref{fig1a.fig} we present a color-color diagram of $\rfUV$
versus $\rfVJ$ for the redshift bins $0<z<1$, $1<z<2$ and $2<z<3.5$.
A clear correlation of $\rfUV$ with $\rfVJ$ is observed at all
redshifts.  The $\rfUV$ color samples the Balmer/4000$\AA$ break.  The
large wavelength range spanned by $\rfUV$ and $\rfVJ$ together is
useful to probe reddening by dust.

To study how the color distribution compares to that in the local
universe, we indicate the colors of galaxies in the low-redshift New
York University Value-Added Galaxy Catalog (NYU\_VAGC, Blanton et
al. 2005) with small grey dots.  The low-z NYU\_VAGC is a sample of
nearly 50000 galaxies at $0.0033<z<0.05$ extracted from the Sloan
Digital Sky Survey (SDSS data release 4, Adelman-McCarthy et
al. 2006).  It is designed to serve as a reliable reference for the
local galaxy population and contains matches to the Two Micron All Sky
Survey Point Source Catalog and Extended Source Catalog (2MASS, Cutri
et al. 2000).  Only the subsample of 20180 sources that are detected
in the 2MASS J-band are plotted in Fig.\ \ref{fig1a.fig}.  This
results effectively in a reduction of the blue peak of the bimodal
$U-V$ distribution.  We only show those galaxies (both for the local
sample and for our sample of HDFS galaxies) with a rest-frame V-band
luminosity $L_V > 5 \times 10^9 L_{\sun}$.  At this luminosity the
distribution of low-z NYU\_VAGC galaxies with SDSS and 2MASS
detections starts falling off.  From the much deeper HDFS imaging the
luminosity cut weeds out low to intermediate redshift dwarf
galaxies.

The same trend of optically red galaxies being red in optical to
near-infrared wavelengths that we found for galaxies up to $z = 3.5$
is observed in the local universe.  However, there are two notable
differences in the color distribution between distant and local
galaxies.  First, a population of luminous high redshift galaxies with
very blue $\rfUV$ and $\rfVJ$ exists without an abundant counterpart
in the local universe.  The 2MASS observations are not deep enough to
probe very blue $V-J$ colors, but we can ascertain that $95\%$ of all
low-z NYU\_VAGC sources with $L_V > 5 \times 10^9 L_{\sun}$ lie in the
range $0.73 < U-V < 2.24$.  About half of the blue galaxies at $z>2$
with $\rfUV < 0.73$ and $L_V > 5 \times 10^9 L_{\sun}$ satisfy the
Lyman-break criterion.  Their stellar populations have been
extensively studied (e.g. Papovich et al. 2001, Shapley et al. 2001,
among many others) and their blue SEDs (see e.g. object $\#242$ and
$\#807$ in Fig.\ \ref{SEDfig1.fig}) are found to be well described by
relatively unobscured star formation.  The rest-frame optical bluing
with increasing redshift of galaxies down to a fixed $L_V$ is
thoroughly discussed by Rudnick et al. (2003).

A second notable difference with respect to the color distribution of
nearby galaxies is present at $\rfUV > 1$, where most local galaxies
reside.  Our sample of HDFS galaxies has a median offset with respect
to the SDSS+2MASS galaxies of $0.22 \pm 0.04$ mag towards redder
$\rfVJ$ at a given $\rfUV$.  Furthermore, the spread in $\rfVJ$ is
larger, extending from colors similar to that of local galaxies to
$\rfVJ$ colors up to a magnitude redder.  The larger spread in $\rfVJ$
colors at a given $\rfUV$ is not caused by photometric uncertainties.
After subtraction in quadrature of the scatter expected from
measurement errors (0.05 mag) we obtain an intrinsic scatter of 0.3
mag, significantly larger than that for SDSS+2MASS galaxies (0.19 mag)
at a $4.5\sigma$ level.

\begin {figure} [tp]
\centering
\resizebox{\hsize}{!}{\plotone{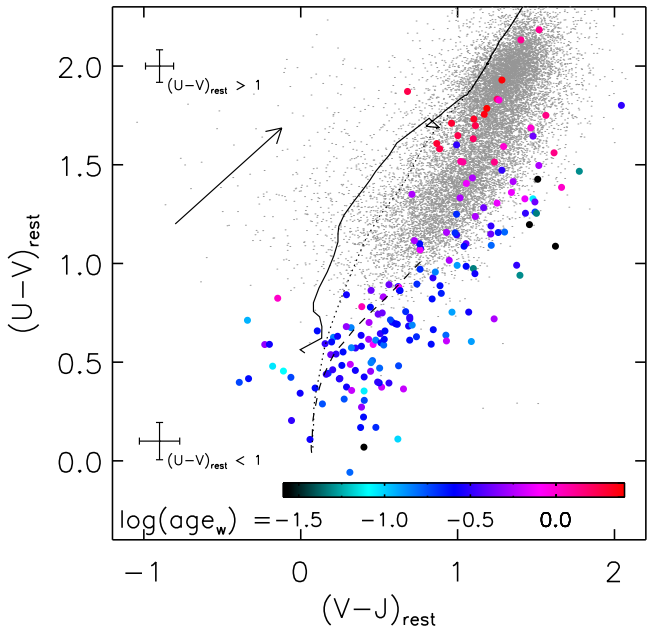}}
\caption{Rest-Frame $U-V$ versus $V-J$ color-color diagram of all
galaxies with $L_V > 5 \times 10^{9}L_{\sun}$.  SDSS+2MASS galaxies
(small grey dots) are plotted as a local reference.  The dust vector
indicates an extinction of $A_V = 1$ mag.  Color evolution tracks of
(unreddened) Bruzual \& Charlot (2003) models are overplotted: a
simple stellar population (SSP, solid line), an exponentially
declining ($\tau_{300}$, dotted line) and constant (CSF, dashed line)
star formation model.  Galaxies are color-coded by best-fit
mass-weighted age.  The tracks show an increase towards redder $U-V$
and slightly redder $V-J$ with age.  At a given $U-V$ color redder
than 1 galaxies that are red in $V-J$ have the youngest best-fit
mass-weighted ages.
\label {fig1b.fig}
}
\end {figure}

In order to understand the nature of galaxies with similar or redder
$\rfVJ$ than the bulk of nearby galaxies, we make use of stellar
population synthesis models by Bruzual \& Charlot (2003; BC03).  In
Fig.\ \ref{fig1b.fig} we draw age tracks for three different dust-free
star formation histories in the $\rfUV$ vs $\rfVJ$ color-color
diagram.
The solid line represents a single stellar population (SSP), the
dashed line a continuous star formation model (CSF) and the dotted
line an exponentially declining star formation model with an e-folding
timescale of 300 Myr ($\tau_{300}$).  All star formation histories
show an evolution to redder $\rfUV$ and $\rfVJ$ with age.  The
$\tau_{300}$ model first has similar colors as a CSF model and
eventually moves to the same region in color space as an evolved SSP,
namely where the red peak of the SDSS bimodal $U-V$ distribution is
located.  In the absence of dust a population with a constant star
formation history only reaches $U-V = 1$ in a Hubble time.

We now investigate how the location in this color plane is related to
stellar populations.  Using the best-fit model parameters (see
\S\ref{SEDmodeling.sec}) we plot the mass-weighted ages for the galaxies
with $L_V > 5 \times 10^9 L_{\sun}$ with color-coding on Fig.\
\ref{fig1b.fig}.  Galaxies with blue optical colors are indeed found
to be young, the median mass-weighted age for galaxies at $\rfUV < 1$
being 250 Myr.  At $\rfUV > 1$ galaxies with a wide range of stellar
ages are found.  The oldest stellar populations show the bluest
$\rfVJ$ colors at a given $\rfUV$.  Over the whole redshift range
galaxies are present that have red optical colors and whose SEDs are
consistent with evolved stellar populations and low dust content.
According to their best-fit model, three of them started forming stars
less than 0.5 Gyr after the big bang and already at $z>2.5$ have star
formation rates less than a percent of the past-averaged value.  We
note that in the Chandra Deep Field South Papovich et al. (2006) find
a number density of passively evolving galaxies at high redshift that
is nearly an order of magnitude lower than in the HDFS, possibly owing
to the fact that the HDFS observations probe to fainter K-band
magnitudes.  The red $\rfVJ$ side of the color distribution is made up
of galaxies that are best fitted by young stellar populations.  Since
the age tracks alone cannot explain the presence of galaxies with such
red SEDs from the optical throughout the near-infrared, we investigate
the role of dust in shaping the galaxy color distribution.

\begin {figure} [tp]
\centering
\resizebox{\hsize}{!}{\plotone{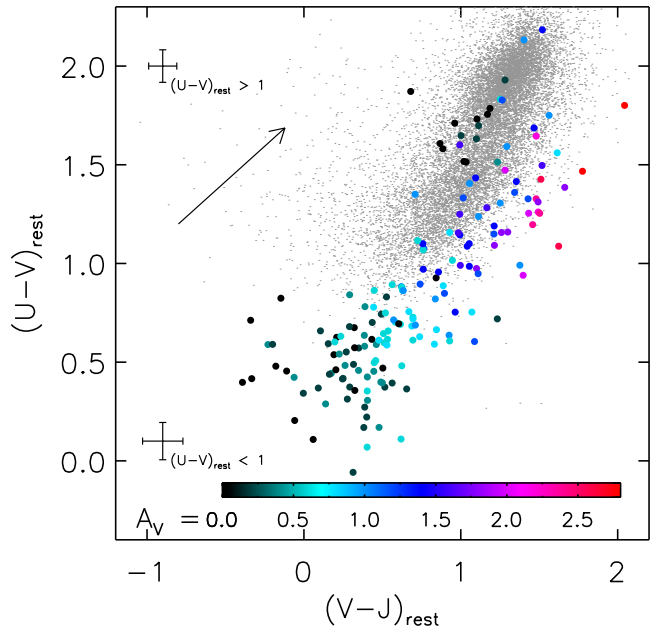}}
\caption{Rest-Frame $U-V$ versus $V-J$ color-color diagram of all
galaxies with $L_V > 5 \times 10^{9}L_{\sun}$.  SDSS+2MASS galaxies
(small grey dots) are plotted as a local reference.  The vector
indicates a dust extinction of $A_V = 1$ mag.  Galaxies are
color-coded by best-fit $A_V$.  The presence of dust moves galaxies to
redder $U-V$ and $V-J$ colors.  Galaxies falling redward in $V-J$ of
the distribution of local galaxies are best described by dusty stellar
populations.
\label {fig1c.fig}
}
\end {figure}

Fig.\ \ref{fig1c.fig} shows again the $\rfUV$ versus $\rfVJ$
color-color diagram, now color-coded by best-fit dust extinction,
expressed in $A_V$.  
The arrow indicates an $A_V$ of 1 magnitude using a Calzetti et
al. (2000) extinction law.  It is immediately apparent that the
optical to near-infrared color-color diagram is a useful diagnostic
for distinguishing stellar populations with various amounts of dust
extinction.  At the bluest $\rfUV$ colors there is little evidence for
dust obscuration.  The degree of dust extinction increases as we move
along the dust vector to redder colors.

Independent constraints on dust-enshrouded activity in distant
galaxies can be derived from MIPS $24\mu m$ imaging (Webb et al. 2006;
Papovich et al. 2006).  The mid-infrared emission is usually thought
to be powered by a dusty starburst in which PAH features are produced
or by an active galactic nucleus (AGN).  Ninety-five \% of the area
with very deep U-to-$8\mu m$ in the HDFS is covered by a 1-hour MIPS
pointing.  We performed the same photometric procedure to reduce
confusion as for the IRAC photometry (see \S\ref{photometry.sec}).
Fluxes were measured within a $6"$ diameter aperture and then scaled
to total using the growthcurve of the $24\mu m$ PSF.

\begin {figure} [tp]
\centering
\resizebox{\hsize}{!}{\plotone{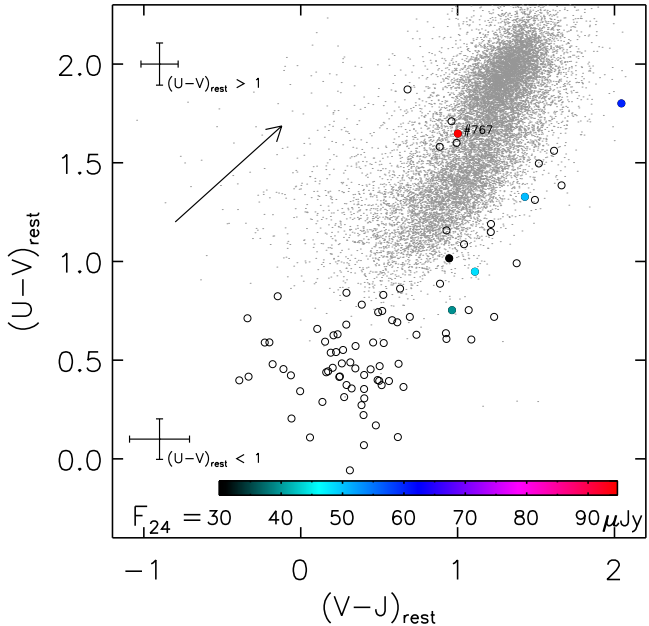}}
\caption{Rest-Frame $U-V$ versus $V-J$ color-color diagram of galaxies
at $1.5 < z < 3$ with $L_V > 5 \times 10^9 L_{\sun}$ with MIPS $24\mu
m$ coverage.  SDSS+2MASS galaxies (small grey dots) are plotted as a
local reference.  Filled circles represent MIPS $24 \mu m$ detections
above a $28 \mu Jy$ ($3\sigma$) threshold.  $\# 767$ is detected at
$24 \mu m$ and has an excess of $8 \mu m$ flux compared to the
best-fitting template SED, suggesting the presence of an obscured AGN.
All other $24 \mu m$ detections lie in the $U-V$, $V-J$ region
populated by galaxies with dusty stellar populations.  Assuming the 24
micron flux originates from PAH emission produced by dust-enshrouded
star formation, the MIPS observations confirm the diagnostic power of
this color combination.
\label {fig1d.fig}
}
\end {figure}

In Fig.\ \ref{fig1d.fig} we plot the $\rfUV$ versus $\rfVJ$
color-color diagram of all objects in the redshift interval
$1.5<z<3.5$ with $L_V > 5 \times 10^9 L_{\sun}$ that are covered by
MIPS (empty circles).  
At these redshifts strong PAH features, if present, move through the
MIPS $24\mu m$ passband.  Six sources have a MIPS $24\mu m$ detection
above $28\mu Jy\ (3\sigma)$.  Their $24\mu m$ flux is indicated by the
colored filled circles.  Object $\# 767$ (red circle) is well detected
with $F_{24\mu m} = 95 \mu Jy$.  As noted by Labb\'{e} et al. (2005)
its SED shows an $8\mu m$ excess with respect to the best-fitting
template.  The combination of $8\mu m$ excess and $24\mu m$ detection
suggests that this galaxy hosts an AGN whose power law SED dominates
throughout the mid-infrared.  All other $24\mu m$ detections are
located in the part of the diagram where our U-to-$8\mu m$ SED
modeling found dusty young populations.  None of the blue relatively
unobscured star-forming galaxies or red evolved galaxies show evidence
of PAH emission from the observed $24\mu m$ flux.  There are various
reasons why not all straforming dusty galaxies have a $24\mu m$
detection.  The density of the UV radiation field exciting the PAHs
may vary among galaxies.  Furthermore, the narrow PAH features with
respect to the width of the $24\mu m$ passband make the $24\mu m$ flux
very sensitive to redshift.  Overall, MIPS observations agree well
with SED modeling and rest-frame optical-NIR color characterization.

\begin {figure} [tp]
\centering
\resizebox{\hsize}{!}{\plotone{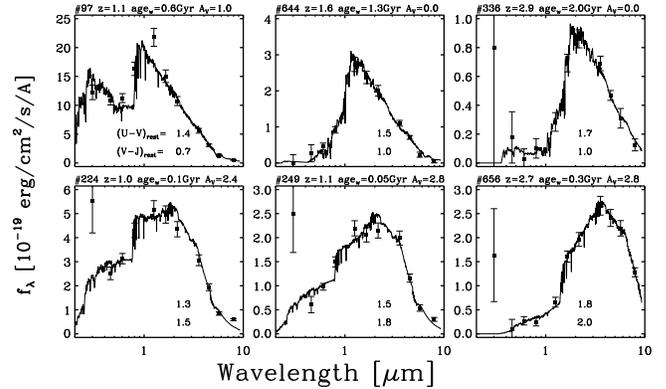}}
\caption{Comparison of galaxies with similar $\rfUV$ color but
different $\rfVJ$ color.  The top row shows the galaxies with blue
$\rfVJ$ colors, and the bottom row shows galaxies with matching
$\rfUV$ color but much redder $\rfVJ$ color.  The systematic
difference in the SEDs of the two rows is striking.  Fits indicate old
bursts of star formation with little dust in the top row, and dusty
young galaxies in the bottom row.  This demonstrates the power of
$\rfVJ$ in separating these classes.  Note that the U-band photometry
for objects $\# 224$ and $\# 249$ deviates by more than $2\sigma$ from
the predicted U-band flux of the best-fit template.
\label {SEDred1.fig}
}
\end {figure}

We conclude that over the whole redshift range from $z=0$ to $z=3.5$ a
trend is visible of galaxies with redder optical colors showing redder
optical to near-infrared colors.  However, at a given optical color, a
spread in optical to near-infrared colors is observed that is larger
than for nearby galaxies.  At $\rfUV > 1$ evolved galaxies with little
dust extinction are found at the bluest $\rfVJ$.  Dusty young
star-forming galaxies occupy the reddest $\rfVJ$ colors.  This is once
more illustrated by the SEDs of galaxies with $\rfUV > 1$ presented in
Fig.\ \ref{SEDred1.fig}.
The upper row shows SEDs of objects at the blue side of the $\rfVJ$
color distribution.  The panels below show SEDs of galaxies matched in
$\rfUV$, but with comparatively redder $\rfVJ$ colors.  The latter
galaxies have comparatively younger ages and a larger dust content.
Since this distinction could not be made on the basis of $\rfUV$ color
alone, the addition of IRAC $3.6 - 8 \mu m$ photometry to our U-to-K
SEDs proves very valuable for the understanding of stellar populations
at high redshift.

We verified that no substantial changes occur to the rest-frame
optical-to-NIR color distribution and its interpretation in terms of
age and dust content of the galaxies when we derive photometric
redshifts by running HYPERZ with redshift as free parameter instead of
using the algorithm developed by Rudnick et al. (2003, see
\S\ref{sample.sec}).

\section {Constraints on stellar population properties at $2<z<3.5$: age, dust and mass}
\label{constraints.sec}
We now proceed to analyse in more detail the constraints that IRAC
places on the stellar populations of the subsample of galaxies at $2 <
z < 3.5$ (75 galaxies).  In particular we will focus on stellar mass,
which likely plays a key role in galaxy evolution at all redshifts
(e.g. Kauffmann et al. 2003; Bundy et al. 2005; Drory et al. 2005;
Rudnick et al. 2006).  Fortunately, estimates of stellar mass from
modeling the broad-band SEDs are generally more robust than estimates
of dust content and stellar age (Bell\& de Jong 2001; Shapley et
al. 2001; Papovich et al. 2001; F\"{o}rster Schreiber et al. 2004).
Nevertheless, translating colors to mass-to-light ratios and
subsequently stellar masses requires a good understanding of the
effects of age and dust.

\subsection {Predictions from stellar population synthesis models}
\label{constr_model.sec}

\subsubsection {Wavelength dependence: optical vs near-infrared}
\label{constr_model_wav.sec}
\begin {figure} [bp]
\centering
\resizebox{\hsize}{!}{\plotone{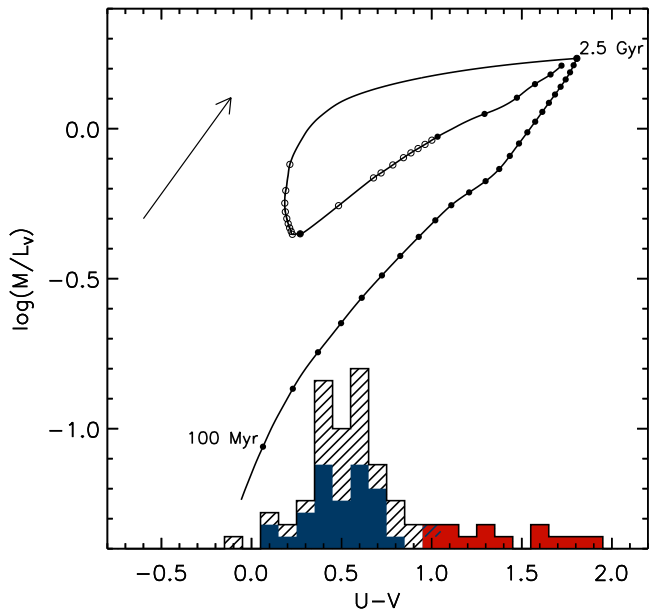}}
\caption{Evolutionary track of a 2-component stellar population in the
$M/L_V$ versus $U-V$ plane.  Filled circles mark age steps of 100 Myr.
Open circles represent 10 Myr age steps.  The dust vector, indicating
an extinction of $A_V = 1$ mag, lies parallel to the age track.  The
histogram represents the color distribution of galaxies at $2 < z <
3.5$, with DRGs highlighted in red, LBGs in blue.  The age track
starts as an exponentially declining star formation model ($\tau =
300$ Myr, BC03).  At 2.5 Gyr a new burst of star formation is
introduced, lasting 100 Myr and contributing 20\% to the mass.
Translating $U-V$ into $M/L_V$ assuming 1-component models can lead to
underestimates of $M/L_V$ and thus stellar mass.  The possible
underestimate is largest for blue galaxies.
\label {fig2a.fig}
}
\end {figure}

In its simplest form, the stellar mass of a galaxy can be estimated
from one color (see e.g. Bell et al. 2001).  To illustrate this
process, we present the evolutionary track of a dust-free BC03 model
in a $M/L_V$ versus $U-V$ diagram (Fig.\ \ref{fig2a.fig}).  
Up to 2.5 Gyr after the onset of star formation (upper right corner of
Fig.\ \ref{fig2a.fig}) the track represents a 1-component population
with a star formation history that is exponentially declining, with an
e-folding timescale of 300 Myr.  For most of the galaxies in our
sample this was the best-fitting star formation history.  For more
extreme star formation histories such as an SSP or CSF the process of
estimating $M/L$ values follows similar arguments.  The filled circles
on Fig.\ \ref{fig2a.fig} represent age steps of 100 Myr.  As the
stellar population ages, its V-band luminosity fades with only a small
decrease in stellar mass from mass loss, moving the galaxy up in
$M/L_V$.  Simultaneously the $U-V$ color reddens as the hot early-type
stars with short lifetimes die.  The dust vector indicating a
reddening of $A_V = 1$ mag runs parallel to the age track of the
1-component model.  Ironically, the mass estimate benefits greatly
from this degeneracy between age and dust in the optical.  Under the
assumption of a monotonic star formation history $\rfUV$ can uniquely
be translated to $M/L_V$, regardless of the precise role of dust or
age.  Only a normalization with $L_V$ is needed to derive the stellar
mass.  A similar relation was used by Rudnick et al. (2003) to
translate the integrated $\rfUV$ color of high redshift galaxies into
a global $M/L_V$ and stellar mass density $\rho_*$.  They found that
the conversion to mass-to-light ratio is more robust from the $\rfUV$
color than from the $(U-B)_{rest}$ or $(B-V)_{rest}$ color.

What if the actual star formation history is more complex?  What
effect does it have on the derived stellar mass?  There is ample
evidence from local fossil records (e.g. Trager et al. 2000; Lancon et
al. 2001; Freeman\& Bland-Hawthorn 2002; F\"{o}rster Schreiber et
al. 2003; Angeretti et al. 2005) and high redshift studies
(e.g. Papovich et al. 2001; Ferguson et al. 2002; Papovich et
al. 2005) that galaxies of various types have complex and diverse star
formation histories, often with multiple or recurrent episodes of
intense star formation .  Such a scenario is also predicted by cold
dark matter models (e.g. Somerville, Primack,\& Faber 2001; Nagamine
et al. 2005; De Lucia et al. 2005).  In order to address this question
qualitatively, we consider the case of a 2-component population.  At
$t=2.5$ Gyr we added a burst of star formation to the $\tau_{300}$
model, lasting 100 Myr and contributing $20\%$ to the mass.  To follow
the evolution of the 2-component population closely, we mark 10 Myr
timesteps with open circles.  Over a timespan of only 10 Myr the
galaxy color shifts by 1.6 mag towards the blue, while the $M/L_V$
value stays well above the 1-component $M/L_V$ corresponding to that
color.  As the newly formed stars grow older, the galaxy moves towards
the upper right corner of the diagram again.  The offset of $M/L_V$
with respect to the 1-component model is a decreasing function of
$U-V$.  This means that if a bursty star formation is mistakenly fit
with a 1-component model the mass and mass-to-light ratio are
underestimated more for blue than for red galaxies, confirming what
Shapley et al. (2005) found for a sample of star-forming galaxies at
$z>2$.

The histogram on the bottom of Fig.\ \ref{fig2a.fig} indicates the
$\rfUV$ color distribution of galaxies in the HDFS at $2<z<3.5$.  The
population of Lyman-break galaxies (LBGs) is marked in blue, Distant
Red Galaxies (DRGs) in red.  The possible underestimate in
mass-to-light ratio and thus mass is largest for blue galaxies, up to
a factor of 3 for $\rfUV = 0.2$, the bluest color reached by this
2-component model.  For DRGs only a modest amount of mass can be
hidden under the glare of a young burst of star formation.  The exact
error that bursts cause depends on the form of the bursty star
formation history (see e.g. Fig. 6 in Rudnick et al. 2003 for a
different example).

\begin {figure} [bp]
\centering
\resizebox{\hsize}{!}{\plotone{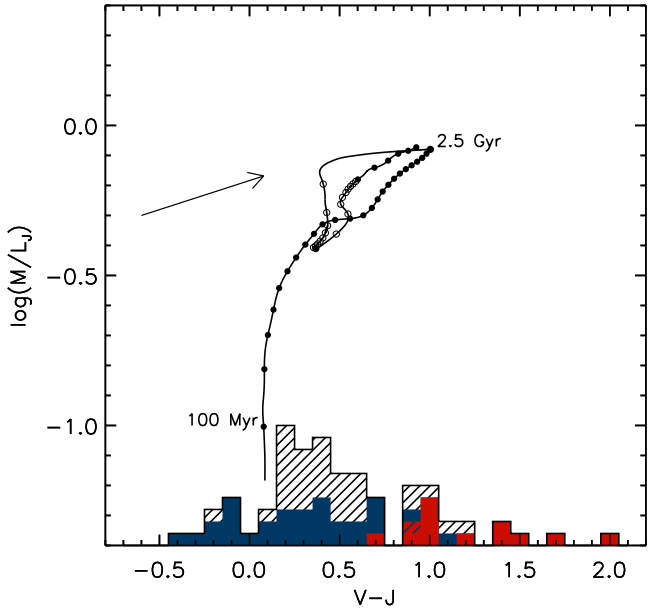}}
\caption{Evolutionary track of a 2-component stellar population in the
$M/L_J$ versus $V-J$ plane.  A $\tau_{300}$ model from BC03 is shown.
At 2.5 Gyr a 100 Myr burst is added , contributing 20\% to the mass.
Age marks represent 100 Myr (filled circles) and 10 Myr (open circles)
respectively.  The histogram shows the color distribution of our
sample at $2 < z < 3.5$, with DRGs in red and LBGs in blue.  For blue
galaxies the $V-J$ color is insensitive to $M/L_J$, further
complicated by the dust vector ($A_V = 1$ mag) that lies nearly
orthogonal to the age track meaning blue galaxies can have a range of
masses for the same $V-J$ color.  On the other hand, the introduction
of a second burst only causes a small offset in $M/L_J$ from the
single-component track, showing that the inclusion of a rest-frame NIR
band reduces the uncertainties in stellar $M/L$ caused by poor
knowledge of the star formation history.
\label {fig2b.fig}
}
\end {figure}

We can now test whether rest-frame near-infrared photometry, as
provided by IRAC, improves the constraints on the SED-based stellar
mass estimates of high redshift galaxies.  Labb\'{e} et al. (2005)
found that the range in $M/L_K$ for DRGs and LBGs together is as large
as a factor 6, meaning that a Spitzer $8\mu m$-selected sample would
be very different from a mass-selected sample.  However, if a similar
relation between mass-to-light ratio and color exists in the
rest-frame near-infrared as in the rest-frame optical, this does not
mean that the stellar mass estimate is uncertain by the same amount (a
factor of 6).  Here we consider whether the mass-to-light ratio can
robustly be derived from a given rest-frame near-infrared color.  We
discuss only the rest-frame J-band but note that the results for
rest-frame K are similar.  In Fig.\ \ref{fig2b.fig} we repeat the same
exercise of drawing a $M/L$ versus color evolutionary track for the
rest-frame near-infrared.  
The burst that we superposed on the $\tau_{300}$ model after 2.5 Gyr
is again contributing $20\%$ to the mass over a period of 100 Myr.
Note that the scale is identical to that of Fig.\ \ref{fig2a.fig}.
The $\rfVJ$ histogram of sources at $2<z<3.5$ is derived from observed
near- to mid-infrared wavelengths.  During the first Gyr, the $V-J$
color hardly changes whereas $M/L_J$ does by a factor of 7.  As an
immediate consequence, the translation of $V-J$ into $M/L_J$ is highly
uncertain for the blue galaxies in our sample and the additional IRAC
observations do not improve the constraints on the mass-to-light
ratio.  The situation is further complicated by the effect of dust.
$V-J$ is a lot more sensitive to dust than $M/L_J$, illustrated by the
dust vector of $A_V = 1$ mag.  The effects of dust and age no longer
conspire to give robust mass estimates at a given $V-J$ color.  At
redder $V-J$ the situation improves as the slope of the age track
flattens.  Here the inclusion of a rest-frame NIR color clearly
reduces the uncertainty in stellar $M/L$ that stems from the poor
knowledge of the star formation history.  The loop towards bluer
colors is a magnitude smaller in size and we see no large offsets in
$M/L$ between the 1- and 2-component modeling.

We have discussed the different behavior of dust and age in simplified
1- and 2-component models and have investigated the improvements
expected from the inclusion of the rest-frame NIR with respect to the
rest-frame optical.  While additional rest-frame NIR data can lead to
better $M/L$ estimates, in particular for redder galaxies ($U-V>1$;
$V-J>0.4$), it is clear that we need to take advantage of the full
U-to-$8\mu m$ SED information to derive reliable estimates of stellar
mass, stellar age and dust content.

\subsubsection {Model dependence: Bruzual \& Charlot vs Maraston}
\label{constr_model_mod.sec}
It is important to note that different stellar population synthesis
models do not paint a consistent picture of evolution in the rest-frame
near-infrared.  To illustrate, we compare Bruzual \& Charlot (2003;
BC03) models to Maraston (2005; M05) models under the same assumption
of Salpeter initial mass function and solar metallicity.

\begin {figure} [bp]
\centering
\resizebox{\hsize}{!}{\plotone{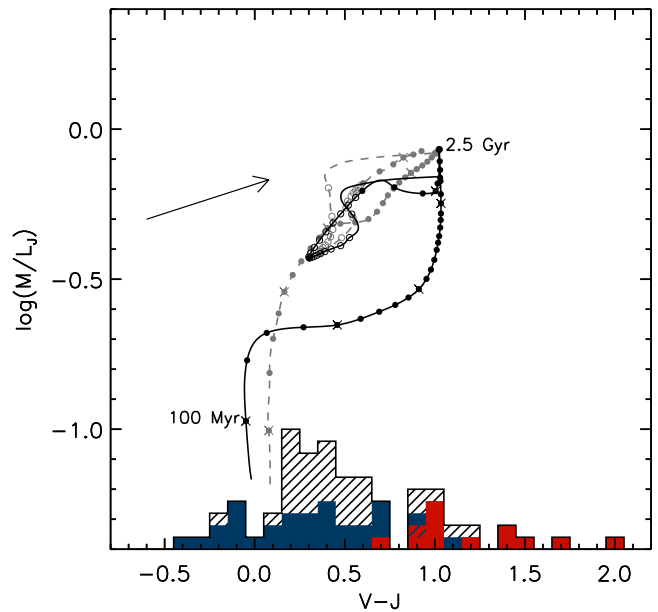}}
\caption{Evolutionary track of 2-component stellar populations in the
$M/L_J$ vs $V-J$ plane based on BC03 (grey dashed line) and Maraston
(2005, M05, black solid line) models.  For ages between 0.2 and 2 Gyr,
the M05 model predicts much lower $M/L_J$ values than the BC03 model.
The underestimate of $M/L_J$ as derived from 1-component modeling is
therefore much more severe for the M05 model than for the BC03 model,
and the inclusion of rest-frame NIR data does not necessarily improve
constraints on stellar $M/L$.
\label {fig2c.fig}
}
\end {figure}

Whereas the age track in a $M/L_V$ versus $U-V$ diagram behaves
similarly for M05 and BC03, the near-infrared evolution of a $\tau_{300}$
model looks very different (see Fig.\ \ref{fig2c.fig}).  
The grey dashed line represents the age track of a BC03 $\tau_{300}$
model with superposed burst at 2.5 Gyr as described in
\S\ref{constr_model_wav.sec}.  In black we overplot the age track of a
2-component model with identical parameters by M05.  In the $0.2 - 2$
Gyr age range the two models look strikingly different.  At the same
$V-J$ color the M05 model predicts $M/L_J$ values that are up to a
factor 2.5 smaller than those of the BC03 model.  The offset between
$M/L_J$ as predicted from 1- and 2-component modeling is also larger
by a similar factor.  The BC03 and M05 models differ in several
aspects: the stellar evolutionary tracks adopted to construct the
isochrones, the synthesis technique and the treatment of the thermally
pulsating Asymptotic Giant Branch (TP-AGB) phase.  The Padova stellar
tracks (Fagotto et al. 1994) used by BC03 include a certain amount of
convective-core overshooting whereas the Frascati tracks (Cassisi et
al. 1997) do not.  The two stellar evolutionary models also differ for
the temperature distribution of the red giant branch phase.  The
higher near-infrared luminosity originates mainly from a different
implementation of the Thermally Pulsating Asymptotic Giant Branch
(TP-AGB) phase (M05).  Following the fuel consumption approach, M05
finds that this phase in stellar evolution has a substantial impact on
the near-infrared luminosity at ages between $0.2$ and 2 Gyr.  BC03
follow the isochrone synthesis approach, characterizing properties of
the stellar population per mass bin.  The latter method leads to
smaller luminosity contributions by TP-AGB stars.  We refer to recent
studies from Maraston et al. (2005), van der Wel et al. (2006) and
Maraston et al. (2006) for discussions of the model differences in
greater detail.

For our purpose it is sufficient to state that a given $V-J$ color
corresponds to younger ages, lower mass-to-light ratios and thus lower
masses for the M05 model than for the BC03 model.  Most importantly,
we note that for M05 models inclusion of NIR data does not reduce
stellar mass uncertainties caused by the unknown star formation
history.

\subsection {Constraints on mass, dust and age from modeling our observed galaxies}
\label{constr_obs.sec}

\subsubsection {Wavelength dependence: optical vs near-infrared}
\label{constr_obs_wav.sec}
Having investigated the qualitative relationship between $M/L$ and the
rest-frame optical-to-NIR color in \S\ref{constr_model_wav.sec}, we now
quantify the effect of inclusion of IRAC MIR photometry on the stellar
population constraints of galaxies at $2<z<3.5$.  Our goal is to
investigate whether and how the addition of IRAC imaging changes our
best estimate of the stellar population properties and their
confidence intervals.

First we compare the distribution of stellar mass, dust content and
mass-weighted stellar age as fit with or without IRAC.  
\begin {figure*} [tp]
\centering
\resizebox{\hsize}{!}{\plotone{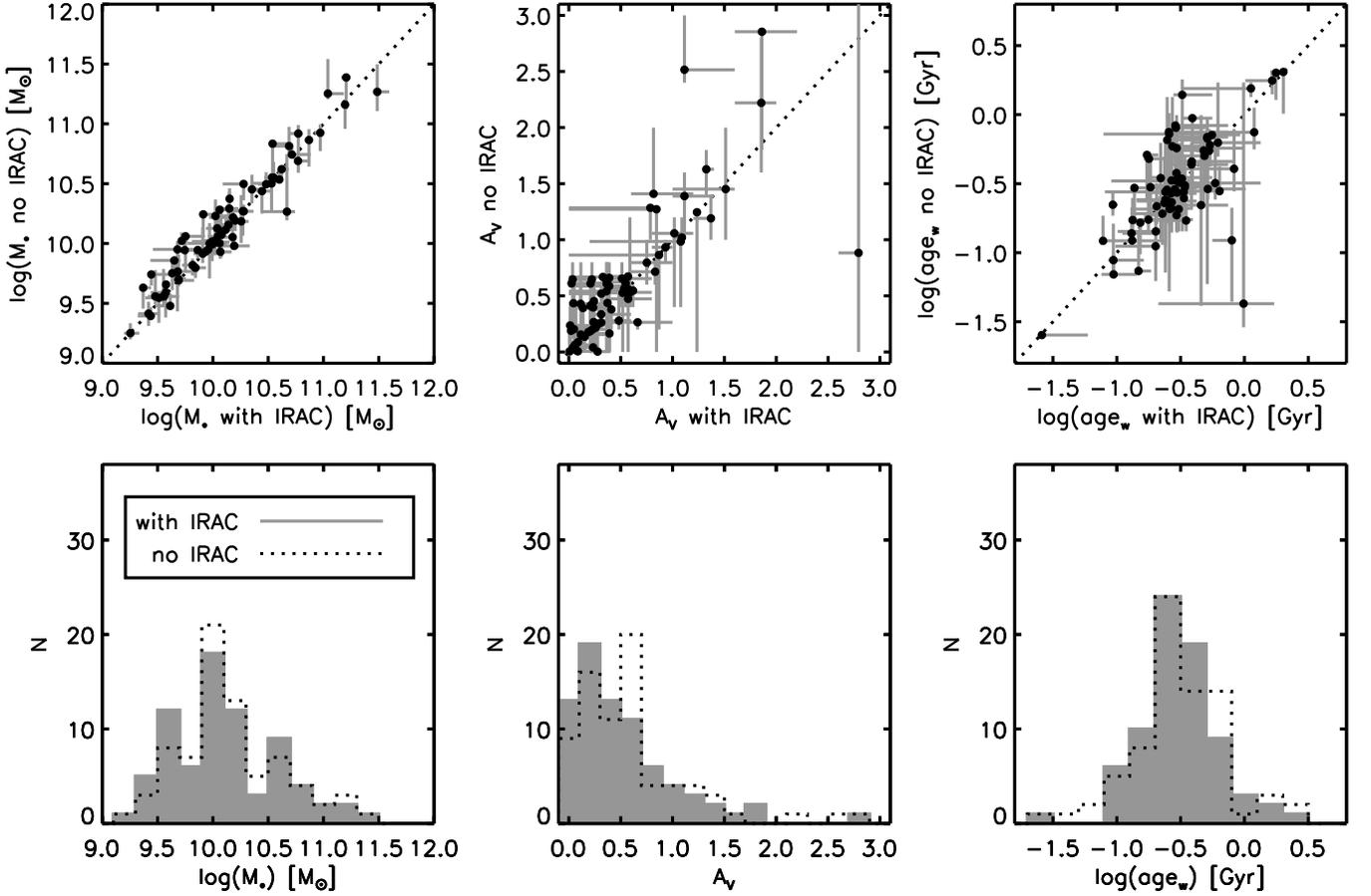}}
\caption{{\it Upper row:} Comparison of best-fit stellar masses, dust
extinctions and mass-weighted ages for galaxies at $2 < z < 3.5$ when
fit with IRAC photometry or without.  The error bars are based on
Monte Carlo simulations given the photometric errors.  {\it Lower
row:} Corresponding histograms with IRAC photometry (filled) or
without (dashed).  No significant change in the overall distributions
is observed, but the best-fit properties of individual galaxies may
change substantially.
\label {fig3.fig}
}
\end {figure*}
The upper row of Fig.\ \ref{fig3.fig} shows a direct comparison of the
inferred model parameters with or without IRAC photometry for all
galaxies at $2<z<3.5$.  The filled histogram in the lower row of Fig.\
\ref{fig3.fig} shows the distribution of mass, dust extinction and age
derived from the full U-to-$8\mu m$ SED.  The dotted line indicates
the distribution of best-fitting parameters from modeling the U-to-K
photometry.  Both the median and the width of the distribution stays
the same for all three parameters.  Defining the difference between
mass, mass-weighted age, and $A_V$ as $\Delta \log (M) = \log (M_{with
IRAC}) - \log (M_{no IRAC})$, $\Delta A_V = A_{V, with IRAC} - A_{V,
no IRAC}$, and $\Delta \log (age_w) = \log (age_{w, with IRAC}) - \log
(age_{w, no IRAC})$ we find a median and normalized median absolute
deviation (equal to the rms for a gaussian distribution) $(\hat{x},
\sigma_{NMAD}(x))$ of $(-0.007 \pm 0.009,\ 0.07)$, $(0.00 \pm 0.03,\
0.30)$, and $(0.00 \pm 0.02,\ 0.16)$ respectively.  The average and
standard deviation $(\langle x \rangle; \sigma(x))$ of $\Delta \log
(M)$, $\Delta A_V$ and $\Delta \log (age_w)$ are $(-0.04 \pm 0.02;
0.13)$, $(-0.08 \pm 0.04; 0.36)$ and $(-0.02 \pm 0.03; 0.28)$
respectively.  Thus the differences for the galaxy sample as a whole
after including IRAC are very small.  The results for stellar mass are
similar to what Shapley et al. (2005) found for a more specific sample
of optically selected star-forming galaxies at $z \sim 2$.

Having determined that the overall distribution of best-fit age, dust
content and stellar mass does not change after including IRAC, the
question remains whether IRAC helps to improve the constraints on the
stellar population characeristics for individual galaxies.  We address
this question using the measure $\sigma_{no IRAC} / \sigma_{with
IRAC}$, defined as the ratio of confidence intervals without and with
IRAC.  The $1\sigma$ confidence intervals, representing random
uncertainties propagating from photometric errors, are derived from
Monte Carlo simulations.  For each galaxy SED we create 100 mock SEDs
where the flux-point in each band is randomly drawn from a Gaussian
with the measured flux as the mean, and its error as the standard
deviation.  Next, each SED was fit with the same fitting procedure as
the observed version.  As we want to isolate the effect of including
IRAC observations on the confidence intervals we fix the redshift to
$z_{phot}$ (or $z_{spec}$ where available).  In calculating
$\sigma_{no IRAC} / \sigma_{with IRAC}$ we measure the confidence
interval in log-space for stellar mass and mass-weighted age and in
magnitude for $A_V$.  Furthermore we set a lower limit to the
confidence intervals to account for the discreteness of our models,
i.e. age and $A_V$ steps.

\begin {figure*} [tp]
\centering
\resizebox{\hsize}{!}{\plotone{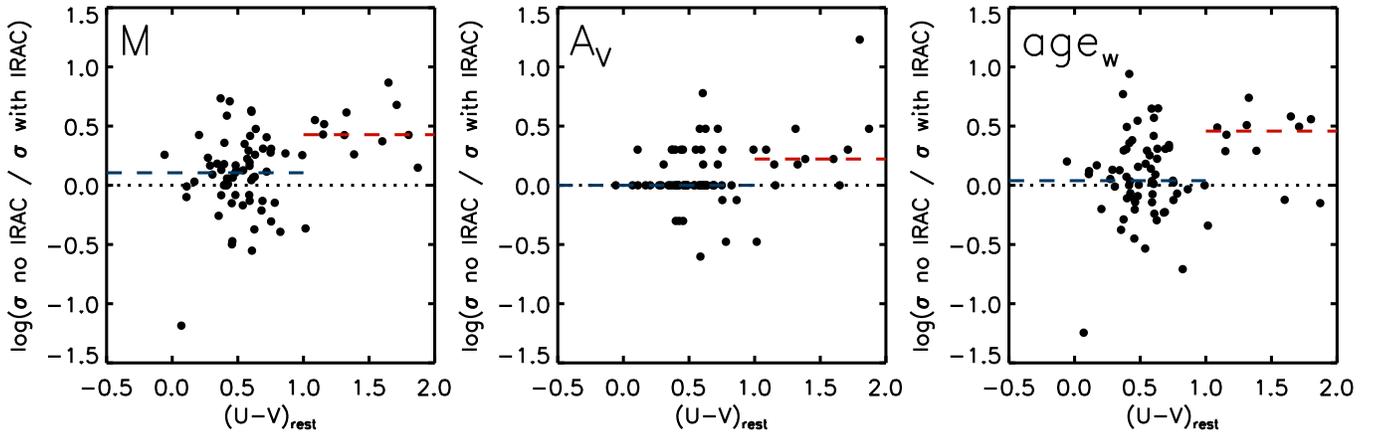}}
\caption{Tightening of the confidence interval around best-fit stellar
mass, age and dust extinction as a function of rest-frame $U-V$ color
for galaxies at $2 < z < 3.5$.  The median improvement after including
the IRAC photometry is a factor 2.7 for red galaxies (dashed red
line), significantly larger than the factor 1.3 for blue galaxies
(dashed blue line).  A similar color-dependence is found for
constraints on age and dust extinction.
\label {fig4.fig}
}
\end {figure*}

Fig.\ \ref{fig4.fig} shows the values of $\sigma_{no IRAC} /
\sigma_{with IRAC}$ for mass, age and dust content as a function of
rest-frame $\rfUV$ color for the galaxies at $2<z<3.5$.  
We divide the sample into a blue and red bin and indicate the median
reduction of confidence intervals for each bin with dashed lines.  The
separation between blue and red is chosen to be $\rfUV = 1$,
corresponding to the oberved $(J-K)_{Vega}>2.3$ color cut for Distant
Red Galaxies at the median redshift of our sample $z=2.66$.  We find
that the typical improvement of confidence intervals is dependent on
galaxy color for all considered stellar population parameters.  For
red galaxies, the reduction amounts to a median factor of 2.7, 1.7 and
2.9 in the case of stellar mass, $A_V$ and age respectively.  For blue
galaxies the reduction of the mass confidence interval is only a
factor 1.3, though with a large scatter, while for $A_V$ and age no
median reduction is found.  With the color tracks of the stellar
population models in mind (see Fig.\ \ref{fig2a.fig}-\ref{fig2b.fig})
this color dependence should come as no surprise.  We demonstrated in
\S\ref{constr_model_wav.sec} that for blue galaxies optical to
near-infrared colors are degenerate with the mass-to-light ratio.
Hence, the IRAC bands of blue galaxies contribute little information
about their mass.

For a sample of (generally blue) optically selected star-forming
galaxies at $z \sim 2$ Shapley et al. (2005) found a reduction in
stellar mass uncertainties by a factor $1.5-2$ due to the addition of
IRAC photometry, which seems like a contradiction.  However, the
distribution of observed $R-K$ color of their galaxies extends towards
redder colors than the Lyman-break galaxies (LBGs) in our sample,
which may partly explain the larger improvement than we find for blue
LBGs.  Another important difference is that Shapley et al. (2005)
lacked J and H images and hence did not probe rest-frame $U-B$ or
$U-V$ for their galaxies.  It is possible that the lack of the near
infrared J and H bands in Shapley et al. (2005) is the main reason for
the discrepancy.  We simulated this effect by omitting J and H and
repeating the Monte Carlo simulations with and without IRAC.  The
median reduction of the $1\sigma$ mass confidence interval now
increases to a factor 1.9 when including IRAC.

We conclude that, in the presence of very deep observed J, H and K
photometry, inclusion of mid-infrared data places little extra
constraints on the stellar populations of blue galaxies.  However, for
galaxies redder than $\rfUV = 1$, IRAC reduces the confidence interval
by a substantial factor $2.5-3$.

\begin {figure*} [htp]
\centering
\resizebox{\hsize}{!}{\plotone{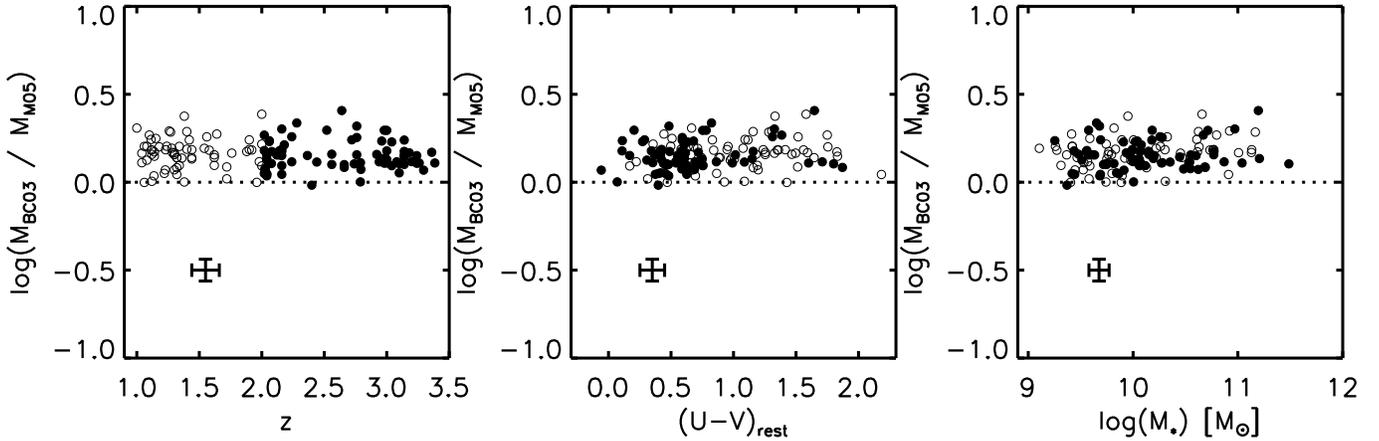}}
\caption{Difference between best-fit stellar mass as derived from BC03
and M05 models as a function of redshift, $\rfUV$ color and BC03
stellar mass for galaxies with $L_V > 5 \times 10^9 L_{\sun}$ at
$1<z<2$ (open symbols) and $2<z<3.5$ (filled symbols).  The stellar
masses derived from BC03 models are systematically higher than those
derived from M05 models by a factor 1.4.  The scatter in
$\log(M_{BC03} / M_{M05})$ is 0.1 in dex.  No significant dependence
of $\log(M_{BC03} / M_{M05})$ on redshift, $\rfUV$ color or stellar
mass is found.  The bias introduced by the choice of stellar
population synthesis model amounts to a maximum of 15\% over the whole
$\rfUV$ color or stellar mass range of our sample.
\label {BC03M05depend.fig}
}
\end {figure*}
\subsubsection {Model dependence: Bruzual \& Charlot vs Maraston}
\label{constr_obs_mod.sec}
In \S\ref{constr_model_mod.sec} we pointed out strong differences in
the rest-frame optical-to-NIR colors between the BC03 and M05 models.
In this paragraph we quantify how our results change if we use M05
models.  The median and normalized median absolute deviation, average
and standard deviation of the differences between BC03 fits and M05
fits ($\Delta \log (M)_{BC03\_M05}$, $\Delta A_{V, BC03\_M05}$ and
$\Delta \log (age_w)_{BC03\_M05}$) are summarized in Table\
\ref{BCvsMar.tab}.
\begin{deluxetable*}{lrrrr}
\tablecolumns{5}
\tablewidth{0pc}
\tablecaption{Differences between stellar population properties derived from BC03 and M05 \label{BCvsMar.tab}
}
\tablehead{
\colhead{} & \colhead{median($\Delta_{BC03\_M05}$)} & \colhead{$\sigma_{NMAD}(\Delta_{BC03\_M05})$} & mean($\Delta_{BC03\_M05}$) & $\sigma (\Delta_{BC03\_M05})$ \\
}
\startdata
$\log (M_*)$    &   $0.14 \pm 0.01$   &  0.06  &  $0.15 \pm 0.01$   &  0.08 \\
$A_V$           &   $-0.20 \pm 0.00$  &  0.00  &  $-0.18 \pm 0.02$  &  0.19 \\
$\log (Age_w)$  &   $0.29 \pm 0.02$   &  0.17  &  $0.34 \pm 0.02$   &  0.18
\enddata
\end{deluxetable*}
As expected, the BC03 models predict older ages and thus higher
stellar masses than the M05 models for our $z = 2 - 3.5$ galaxies.
The estimated mass for the M05 models is systematically lower by a
factor 1.4.  Maraston et al. (2006) found a similar discrepancy for a
sample of 7 galaxies in the Hubble Ultra Deep Field that satisfy the
BzK criterion (Daddi et al. 2004) for $z>1.4$ passively evolving
galaxies.  Apart from a systematic shift a scatter of 0.1 in dex
is found in $\Delta \log (M)_{BC03\_M05}$, meaning the choice of stellar
population synthesis model introduces a considerable systematic
uncertainty.  

It is of great importance to test whether $\Delta \log (M)_{BC03\_M05}
= \log(M_{BC03}) - \log(M_{M05})$ correlates with redshift, color or
stellar mass, since such dependencies, if present, could bias studies
of galaxy evolution or trends with mass.  In Fig.\
\ref{BC03M05depend.fig} we plot $\Delta \log (M)_{BC03\_M05}$ versus
redshift, $\rfUV$ color, and stellar mass (the latter derived from
BC03 models).
We show galaxies with $L_V > 5 \times 10^9 L_{\sun}$ at $1<z<2$ (open
symbols) and at $2<z<3.5$ (filled symbols); no evidence for a redshift
dependence is found.  For the $\rfUV$ (middle) and stellar mass
(right) panel, the p-values for statistical significance from the
Spearman rank order correlation test are also larger than 0.05,
meaning no significant correlation is found.  Fitting a line to the
points in the $\Delta \log (M)_{BC03\_M05}$ versus $\rfUV$ diagram, a
difference of 0.06 dex in $\Delta \log (M)_{BC03\_M05}$ is found over
the 2 mag range in $\rfUV$ color spanned by the galaxies in our
sample.  Even if a trend of increasing $\Delta \log (M)_{BC03\_M05}$
with redder $\rfUV$ color is real, it only introduces a small bias of
the order of 15\%.  A similar conclusion can be drawn for the
dependence on stellar mass.

\subsubsection {Metallicity dependence}
\label{metal.sec}
We test how variations from solar metallicity affect the estimates of
stellar mass, mass-weighted age and dust extinction.  We study the
effect of a different metal abundance by fitting BC03 templates with
metallicity $Z=0.2\ Z_{\sun}$ to the observed SEDs, leaving the
extinction law to Calzetti et al. (2000).  NIR spectroscopy of DRGs
(van Dokkum et al. 2004) and LBGs (Erb et al. 2006a) indicates that a
range of $Z=0.2-1\ Z_{\sun}$ is appropriate for galaxies at $2<z<3.5$.
Furthermore, at metallicities below $Z=0.2\ Z_{\sun}$ the tracks and
spectral libraries used to build the BC03 templates become more
uncertain by lack of observational constraints.  Decreasing the
metallicity from $Z=Z_{\sun}$ to $Z=0.2\ Z_{\sun}$ lowers the
estimated stellar masses of galaxies at $2<z<3.5$ by 0.1 dex, leads to
a mass-weighted age that is typically lower by 0.2 dex, and is
compensated by an average increase in $A_V$ of 0.2 mag.  The fact that
age estimates are more strongly affected than estimates of stellar
mass when changing the assumed metallicity was demonstrated in detail
by Worthey (1994).  While absolute values of ages and dust extinctions
may be biased as just described, the relative age and dust trends
within the galaxy population as discussed in
\S\ref{color_distribution.sec} based on the standard SED modeling (see
\S\ref{SEDmodeling.sec}) are robust.

\subsubsection {Dependence on extinction law}
\label{extinction.sec}
The Calzetti et al. (2000) extinction law was empirically derived from
observations of local starburst galaxies.  We quantify the variations
in stellar population properties due to the adopted extinction law by
comparing our modeling results with a Calzetti et al. (2000) law to
those obtained with reddening laws from Fitzpatrick (1986) for the
Large Magellanic Cloud (LMC) and Pr\'{e}vot et al. (1984) for the
Small Magellanic Cloud (SMC), leaving the metallicity to solar.
Stellar masses, mass-weighted ages and $A_V$ values of galaxies at
$2<z<3.5$ derived with the LMC law models are similar to those
obtained with the Calzetti et al. (2000) law.  The SMC law, which
rises more steeply towards shorter wavelengths in the near-UV, gives
similar mass estimates, $A_V$ values that are on average smaller by
0.3 mag and mass-weighted stellar ages that are older by 0.23 dex, with
the ages of the oldest galaxies being limited by the age of the
universe constraint.  As for metallicity, we conclude that using a
different extinction law has a larger impact on the age estimates than
on estimates of stellar mass.

\section {Stellar mass - optical color relation}
\label {mass_rfUV.sec}

\begin {figure} [bp]
\centering
\resizebox{\hsize}{!}{\plotone{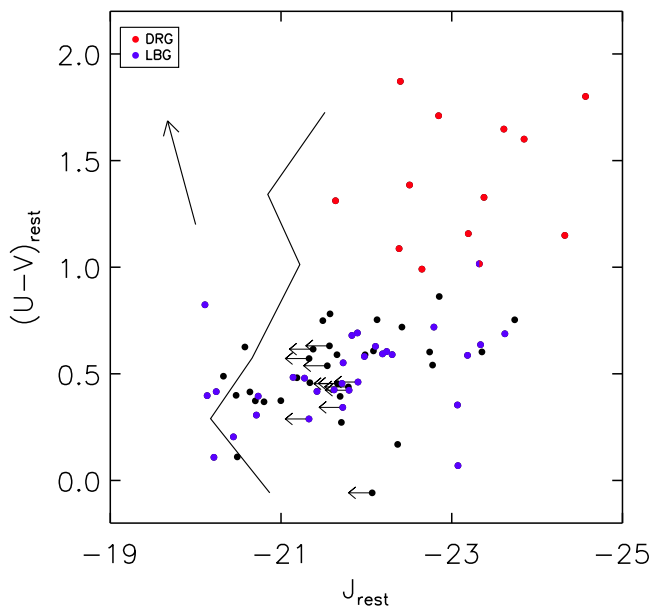}}
\caption{Rest-Frame $U-V$ color versus absolute J magnitude for
galaxies at $2 < z < 3.5$.  Lyman-break galaxies are plotted in blue.
DRGs (red symbols) populate the red side of the $U-V$ color
distribution.  Black symbols denote those objects that do not meet
either criteria.  The solid line marks the K-band selection of our
sample.  The dust vector indicates an extinction of $A_V = 1$ mag.
The most luminous galaxies in the rest-frame near-infrared have redder
rest-frame optical colors than fainter galaxies.
\label {fig5a.fig}
}
\end {figure}

In this section we study the relation between the rest-frame optical
color of high redshift galaxies and their stellar mass.  We start with
a model-independent approach, plotting rest-frame $\rfUV$ versus
rest-frame $J_{rest}$ magnitude for all galaxies at $2<z<3.5$ in Fig.\
\ref{fig5a.fig}.  
The emission of low mass long-lived stars that make up the bulk of the
mass in a galaxy peaks in the rest-frame near-infrared.  $J_{rest}$ is
therefore expected to be a reasonably good tracer of stellar mass.
The galaxies that satisfy the DRG selection criterion (red symbols)
are found at redder $\rfUV$ than the Lyman-break galaxies (blue
symbols).  The reddest $\rfUV$ colors are found at the brightest
$J_{rest}$ magnitudes.  Note however that the observed trend is
partially driven by the K-band selection of our sample.  The line on
Fig.\ \ref{fig5a.fig} indicates at which magnitude a galaxy with
identical colors to our observed galaxies would fall out of the
sample.  Even if we only consider galaxies brighter than the limiting
$J_{rest} = -21.5$ to which we are complete over the whole $\rfUV$
color range, we find that galaxies redder than $\rfUV = 1$ are 1 mag
brighter than galaxies with $\rfUV < 1$, significant at the $3\sigma$
level.  Studying a sample without color bias (as advocated by van
Dokkum et al. 2006) proves crucial to pick up the trend of $\rfUV$
with $J_{rest}$.  We note that Meurer et al. (1999) found that LBGs
with higher rest-frame UV luminosities tend to have redder rest-frame
UV colors, illustrating that, while trends of color with luminosity
are most notable in samples without color bias, they are still present
in at least some color selected samples.

\begin {figure} [tp]
\centering
\resizebox{\hsize}{!}{\plotone{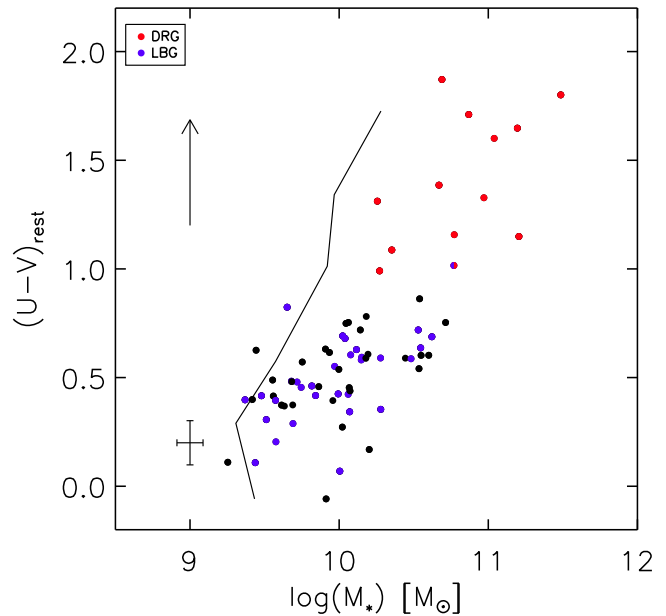}}
\caption{Rest-Frame $U-V$ color versus stellar mass for galaxies at $2
< z < 3.5$.  DRGs are marked in red, LBGs in blue.  The K-band
selection of our sample is indicated by the solid line.  The dust
vector indicates an extinction of $A_V = 1$ mag.  Low-mass galaxies
with red colors might exist, but would not enter the sample.  The most
massive galaxies have redder $U-V$ colors than less massive galaxies;
notice the striking absence of massive blue galaxies.
\label {fig5b.fig}
}
\end {figure}

If $J_{rest}$ is a reasonable tracer of stellar mass, we expect to see
a similar or stronger trend of $\rfUV$ with the stellar mass.  This is
shown in Fig.\ \ref{fig5b.fig}.  
The plotted mass is derived from 1-component SED modeling of the
U-to-$8\mu m$ SED as described in \S\ref{SEDmodeling.sec}.  The
typical error bar is indicated in the lower left corner.  The depth of
our K detection band allows us to probe stellar masses from $3 \times
10^{11}\ M_{\sun}$ down to $2 \times 10^9\ M_{\sun}$.  A correlation
of $\rfUV$ with stellar mass is clearly visible.  The most massive
galaxies have a red optical color.  Lyman-break galaxies and other
blue galaxies at high redshift contain typically 5 times less stellar
mass than the DRGs in our sample.  Again the K-band selection of our
sample (solid line) limits our ability to detect faint red galaxies.
Therefore we can not exclude the presence of low-mass red galaxies.
The lack of massive blue galaxies seems to be real.  Rigopoulou et
al. (2006) find a co-moving density of $\Phi = (1.6 \pm 0.5) \times
10^{-5}\ Mpc^{-3}$ for LBGs with $M > 10^{11} M_{\sun}$ at an average
redshift $\langle z \rangle \simeq 3$, consistent with the absence of
such massive but rare LBGs in our sample.

However, the lack of massive blue galaxies could be an artifact of our
choice of simple star formation histories.  As demonstrated in
\S\ref{constr_model_wav.sec} a severe underestimate of the stellar
mass is possible when the true star formation history is more complex
than that of the modeled 1-component stellar population.  When a young
burst of star formation is superposed on a maximally old population,
its blue light will dominate the $\rfUV$ color and the mass from the
underlying population will be hidden.  In order to constrain the
possible underestimate in mass, we fit 2-component models to our SEDs.
Erb et al. (2006b) describe a procedure to achieve this in two steps,
where first a maximally old population is fit to the K(+IRAC) data and
subsequently a young population is fit to the (primarily UV) residual.
However, this procedure does not guarantee a good fit in the $\chi ^2$
sense.  Instead, we decided to perform a simultaneous fit of both old
and young components.  We constructed template SEDs consisting of a
maximally old single stellar population with a recent burst of star
formation that started 100 Myr ago and lasted till the moment of
observation superposed.  We made templates where the mass fraction
created in the burst is $2^x$ with x going from -6 to 2 in steps of 1.
We assume that the same reddening by dust applies to the old and the
young population, with $A_V$ ranging from 0 to 3 in steps of 0.2.
Without this assumption, one could in principle hide an infinite
amount of mass in an old population as long as an optically thick
medium is shielding it from our sight.  However, such a scenario is
physically implausible.  Since we are interested in an upper limit on
the mass, as opposed to the most likely value, we do not search for
the least-squares solution over all x.  Instead we perform the fit for
every burst fraction and select the highest mass that still has
$\Delta \chi_{red} ^2 = \chi^2_{red,2-component} - \chi ^2_{red,min,
1-component} < 2$.

\begin {figure} [tp]
\centering
\resizebox{\hsize}{!}{\plotone{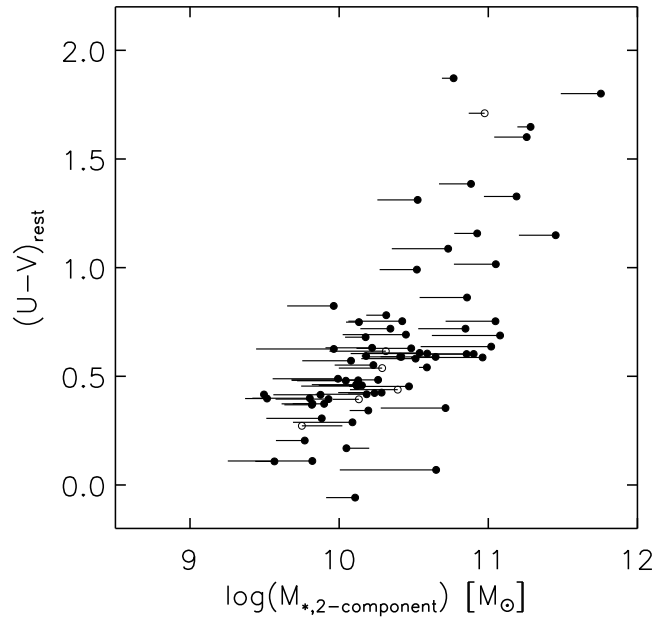}}
\caption{Rest-Frame $U-V$ color versus stellar mass for galaxies at $2
< z < 3.5$ with complex star formation histories.  For each object a
vector starts at the best-fit mass from 1-component SED modeling.  The
vector ends at an upper limit for the stellar mass as obtained by
fitting 2-component models.  The 2-component model is composed of a
maximally old stellar population with a second burst of star formation
during the last 100 Myr superposed.  Empty symbols refer to objects
for which $\chi_{red,2-component}^2 - \chi_{red,1-component}^2 > 2$
for all considered burst fractions.  The mass that is plotted here
corresponds to the burstfraction that gave the lowest $\chi_{red}^2$.
The typical amount of mass that can be hidden under the glare of a
young secondary burst is on average larger for blue than for red
galaxies.  Nevertheless, even allowing for more complex star formation
histories than 1-component models a lack of massive blue galaxies
remains visible.
\label {fig5c.fig}
}
\end {figure}

Fitting the 2-component models to the U-to-$8\mu m$ SEDs of our galaxy
sample at $2 < z < 3.5$, we indeed see that a higher stellar mass is
allowed when more complex star formation histories are adopted (Fig.\
\ref{fig5c.fig}).  
The upper bound on stellar mass that we derive from this particular
2-burst model is in the median a factor 1.7 higher than the
1-component estimate for galaxies redder than $\rfUV = 1$.  For blue
galaxies the median increase is a factor 2.1.  Despite the fact that
more mass can be hidden in blue galaxies, a trend of optical color
with stellar mass remains visible.  We performed a Mann-Whitney U-test
to compare the $\rfUV$ colors of galaxies with different stellar mass.
We conservatively adopted the 1-component stellar mass for galaxies
with $\rfUV > 1$ and the 2-component upper limit for objects with
$\rfUV < 1$.  To avoid selection effects we only consider galaxies
more massive than $M = 10^{10}\ M_{\sun}$.  Dividing them in two mass
bins with an equal number of objects the Mann-Whitney U-test (Walpole
\& Myers 1985) confirms at a $99\%$ significance level that the mean
of the $\rfUV$ distributions differs.  Applying the same 2-component
models to the U-to-K SEDs (omitting IRAC), the median upper mass
estimate increases to a factor 2.3 above the 1-component estimate for
red objects and a factor 3.7 for blue objects.  We conclude that, as
expected from \S\ref{constr_model_wav.sec}, more mass can be hidden in
blue than in red galaxies, but this effect is insufficient to remove
the trend of stellar mass with color.  Furthermore, the amount of mass
that can be hidden is constrained by addition of IRAC photometry.

The color dependence that we derive for the amount of mass that can be
hidden in an underlying old population confirms findings from Shapley
et al. (2005) based on a sample of star-forming galaxies at $z \sim
2$.  The predominance of distant red galaxies at the high-mass end was
illustrated recently by van Dokkum et al. (2006) using a mass-selected
sample of galaxies at $2<z<3$ with $M > 10^{11}\ M_{\sun}$.  Only with
very deep imaging such as that of the HDFS analyzed in this paper it
is possible to probe down to lower masses and prove that the most
massive galaxies have red $\rfUV$ colors compared to lower mass
galaxies.

\section {Summary}
\label{conclusions.sec}
We investigated the rest-frame optical to near-infrared color
distribution of galaxies up to $2<z<3.5$ in the Hubble Deep Field
South.  At all redshifts, galaxies with redder $\rfUV$ tend to have
redder $\rfVJ$, as is the case in the local universe.  At $\rfUV$
colors comparable to that of local galaxies, the color distribution of
distant galaxies extends to redder $\rfVJ$.  At $\rfUV > 1$ the
population of galaxies at the red $\rfVJ$ end is well described by
dust-enshrouded star-forming models, whereas galaxies with $\rfVJ$
similar to that of local galaxies are consistent with old passively
evolving systems.  We conclude that $\rfUV$ alone allows us to isolate
blue relatively unobscured star-forming galaxies, but addition of
$\rfVJ$ is necessary to distinguish young dusty from old passively
evolved systems.  At redshifts above $z=1$, this means IRAC
observations are crucial in understanding the wide variety in stellar
populations.  We note that our analysis is not subject to
uncertainties due to field-to-field variations, but surveys over much
larger areas are needed to study the relative contributions of
galaxies with different stellar populations.  

We analyzed the constraints that IRAC places on stellar mass, stellar
age and dust content of galaxies at $2<z<3.5$.  No evidence is found
for systematic offsets when determining the stellar population
characteristics with or without IRAC.  However, the ratio of
confidence intervals on stellar mass, mass-weighted age and dust
extinction is typically reduced by a factor 2.7, 2.9 and 1.7
respectively for red ($\rfUV > 1$) galaxies.  In general, IRAC does
not provide stronger constraints for blue galaxies ($\rfUV < 1$) when
very deep near-infrared imaging is available (as is the case for the
HDFS).

We caution that, in characterizing the stellar populations using M05
models, we find stellar masses that are typically a factor 1.4 lower
than for BC03 models with a scatter of 0.1 in dex.

A trend of brighter $J_{rest}$ with redder $\rfUV$ is observed for
galaxies at $2<z<3.5$, where the near-infrared luminosity serves as a
(imperfect but model-independent) tracer for stellar mass.  Plotting
$\rfUV$ versus modeled stellar mass, we arrive at a similar
conclusion: the most massive galaxies in our sample have red rest-frame
optical colors.  A possible concern is that this trend with mass is
caused by our simplistic choice of star formation histories.  When we
allow for more complex star formation histories, more mass can be
hidden than in the case of a 1-component stellar population and the
amount depends on the color of the galaxy.  We used 2-component
stellar populations, consisting of a maximally old population with a
young population superposed, to set an upper bound on the stellar mass
present.  Even though relatively more mass can be hidden in blue
galaxies compared to red galaxies, under the assumption of an equal
dust reddening of the young and old component, a trend of $\rfUV$
increasing to redder colors with stellar mass remains visible.






\end {document}